\documentclass[12pt,preprint]{aastex}

\usepackage{color}

\shorttitle{High Latitude Filament Chirality}
\shortauthors{Yeates \& Mackay}

\begin{document}


\title{Chirality of High Latitude Filaments over Solar Cycle 23}


\author{A. R. Yeates}
\affil{Department of Mathematical Sciences, Durham University, Durham, DH1 3LE, UK}
\email{anthony.yeates@durham.ac.uk}

\author{D. H. Mackay}
\affil{School of Mathematics \& Statistics, University of St Andrews, St Andrews, KY16 9SS, UK}
\email{duncan@mcs.st-and.ac.uk}


\begin{abstract}
A non-potential quasi-static evolution model coupling the Sun's photospheric and coronal magnetic fields is applied to the problem of filament chirality at high latitudes.
For the first time, we run a continuous 15 year simulation, using bipolar active regions determined from US National Solar Observatory, Kitt Peak magnetograms between 1996 and 2011. 
Using this simulation, we are able to address the outstanding question of whether magnetic helicity transport from active latitudes can overcome the effect of differential rotation at higher latitudes. Acting alone, differential rotation would produce high latitude filaments with opposite chirality to the majority type in each hemisphere.
We find that differential rotation can indeed lead to opposite chirality at high latitudes, but only for around 5 years of the solar cycle following the polar field reversal.
At other times, including the rising phase, transport of magnetic helicity from lower latitudes overcomes the effect of {\it in situ} differential rotation, producing the majority chirality even on the polar crowns at polar field reversal.
These simulation predictions will allow for future testing of the non-potential coronal model.
The results indicate the importance of long-term memory and helicity transport from active latitudes when modeling the structure and topology of the coronal magnetic field at higher latitudes.
\end{abstract}




\section{Introduction}

Observations such as X-ray sigmoids, flares, or eruptions strongly indicate the presence of substantial electric currents in the low solar corona (below, say, $2.5R_\odot$), which are absent by definition in traditional potential field extrapolations \citep{altschuler1969,schatten1969}. To account for the development and effects of these electric currents, we have introduced a non-potential (hereafter NP) model coupling the Sun's photospheric and coronal magnetic fields \citep{vanballegooijen2000,yeates2008,yeates2010}. The NP model circumvents the problem of non-uniqueness which plagues the extrapolation of nonlinear force-free equilibria from observed photospheric magnetograms \citep{derosa2009} by building up currents in a time-dependent manner. The coronal magnetic field relaxes continually (toward a nonlinear force-free state) in response to shearing by large-scale photospheric motions, and new flux emergence. Previous testing \citep{yeates2008,yeates2010} has shown this model to better approximate the real physics determining the large-scale coronal magnetic field.

\begin{figure}
\includegraphics[width=0.3\textwidth]{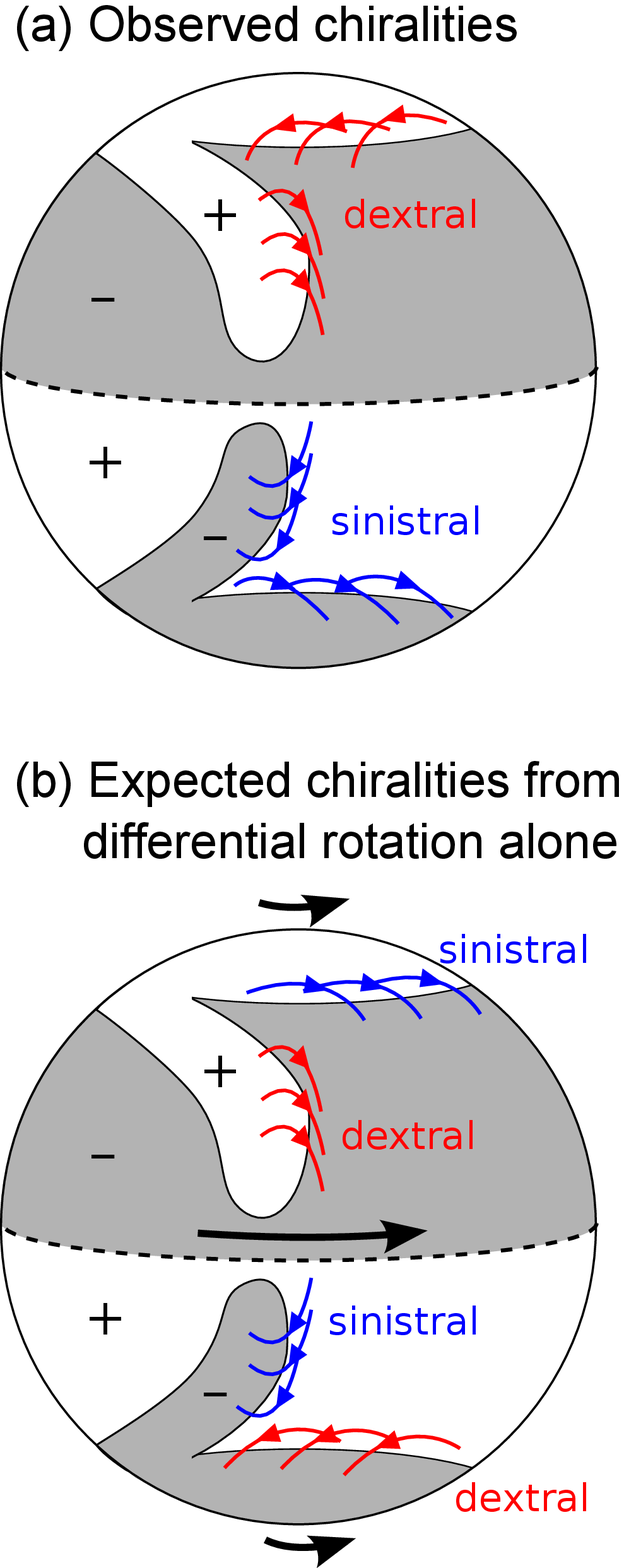}
\caption{Filament magnetic fields are assigned dextral or sinistral chirality according to the angle between their horizontal field and the underlying PIL. Observations indicate the global chirality pattern in (a) \citep[e.g.,][]{leroy1983}, while differential acting alone on an unsheared magnetic arcade would lead to the pattern in (b) \citep{zirker1997}.
See the electronic edition of the Journal for a color version 
of this figure.
\label{fig1}}
\end{figure}

To test the global NP model, \citet{yeates2008} compared the simulated magnetic field direction with the chirality of quiescent solar filaments observed in H$\alpha$ over a 6-month period, finding excellent agreement. Solar filaments are stable structures of (relatively) cool, dense plasma suspended well above the chromosphere \citep{labrosse2010,mackay2010}, and overlying Polarity Inversion Lines (PILs) in the photospheric magnetic field. They are seen in both H$\alpha$ and Extreme Ultraviolet, either against the solar disk or as prominences above the limb. The large-scale magnetic field in filaments is believed to be orientated along the filament axis. Filaments may then be classified as having either dextral or sinistral chirality, according to the direction of this axial field relative to the underlying photospheric polarities (Figure \ref{fig1}). Although a property of the filament magnetic field, \citet{martin1994} showed that the chirality may often be determined from the orientation of barbs in H$\alpha$ images, given only the photospheric polarities on either side of the filament channel. Observations reveal a statistical preference for dextral chirality in the northern hemisphere and sinistral in the south \citep{rust1967,leroy1983,martin1994,pevtsov2003,bernasconi2005}, as indicated in Figure \ref{fig1}(a).

For filaments equatorward of about $60^\circ$ latitude, we were able to use the NP model to explain both the observed pattern and how exceptions can occur. The chirality results from a number of competing effects, depending on the filament location \citep{mackay2001,mackay2005,yeatesthesis,yeates2009}. PILs within individual active regions have chirality consistent with the helicity sign in the active region \citep{rust1994}. But PILs between or outside of active regions can be strongly affected by interactions between multiple regions---this explains many exceptions to the majority hemispheric pattern \citep[supported by the observations of][]{wang2007, romano2009}. Also, over time, differential rotation can strengthen, weaken, or even reverse the chirality at a given PIL \citep{yeatesthesis}.

At higher latitudes, projection effects make it difficult to determine filament chirality through H$\alpha$ barbs \citep[for example, polar crown filaments were omitted by][]{pevtsov2003}. However, several sets of linear polarization measurements in limb prominences indicate the same hemispheric chirality pattern, both at solar minimum \citep{rust1967,leroy1978} and in the rising phase of the solar cycle \citep{leroy1983}, although (to our knowledge) no studies during the declining phase have been published. This finding that the same hemispheric pattern holds at higher latitudes was not reproduced by our previous 6-month simulations for 1999 (in the rising phase of Cycle 23), where the polar crown chirality was found to be opposite to that of lower latitudes.

We attributed this result to the dominance of differential rotation on the East-West oriented polar crown PILs. As shown in Figure \ref{fig1}(b), differential rotation alone leads to the wrong chirality overlying East-West oriented PILs \citep{leroy1978,zirker1997}. However, the chirality of high-latitude PILs in the NP model is a balance between differential rotation and the poleward transport of magnetic helicity from lower latitudes. (We are assuming that there is no net injection of helicity from the unresolved small-scale motions.) The possibility therefore remains that there was simply insufficient time in the 6-month simulations for poleward transport of helicity. In this Letter we extend the simulation to 15 years' duration, simulating both the rising and decaying phases of the solar cycle, to explore the predicted chirality at high latitudes.

\section{Simulation of Cycle 23}

We simulate the evolution of the large-scale photospheric and coronal magnetic fields between 1996 and 2011, using the non-potential (NP) model, which couples surface flux transport to magneto-frictional relaxation in the corona above \citep{vanballegooijen2000,yeates2008,yeates2010}. The large-scale mean magnetic field, ${\bf B}_0=\nabla\times{\bf A}_0$, in the corona is evolved via the induction equation
\begin{equation}
\frac{\partial {\bf A}_0}{\partial t} = {\bf v}_0\times{\bf B}_0 - {\bf E}_0
\end{equation}
where we neglect Ohmic diffusion and the mean electromotive force ${\bf E}_0$ describes the effect of unresolved small-scale fluctuations. Following \citet{vanballegooijen2008} we apply a hyperdiffusion 
\begin{equation}
{\bf E}_0=-\frac{{\bf B}_0}{B_0^2}\nabla\cdot\Big(\eta_4B_0^2\nabla\alpha_0\Big),
\end{equation}
where
\begin{equation}
\alpha_0=\frac{{\bf B}_0\cdot\nabla\times{\bf B}_0}{B_0^2}
\label{eqn:alpha0}
\end{equation}
and we take $\eta_4=10^{11}\,{\rm km}^4{\rm s}^{-1}$. This form of hyperdiffusion preserves magnetic helicity density ${\bf A}_0\cdot{\bf B}_0$ in the volume and describes the tendency of the magnetic field to relax to a state of constant $\alpha_0$ \citep{boozer1986,bhattacharjee1986}, 
although such a state is never reached in the global simulation. We have repeated the simulation with a constant diffusivity (${\bf E}_0=-\eta{\bf j}_0$) with no change to the chirality results.
The velocity is determined by the magneto-frictional technique \citep{yang1986} as
\begin{equation}
{\bf v}_0=\frac{1}{\nu}\frac{{\bf j}_0\times{\bf B}_0}{B_0^2} + v_{\rm out}(r)\hat{\bf r},
\end{equation}
where the first term enforces relaxation toward a force-free equilibrium, and the second term is a radial outflow imposed only near the outer boundary ($r=2.5R_\odot$) to represent the effect of the solar wind radially distending magnetic field lines \citep{mackay2006}.

To run the simulation continuously over 15 years, we use a non-uniform spherical grid whose resolution reduces from $192$ grid points in longitude at the equator to only $12$ at the polar grid boundaries ($\pm89.5^\circ$ latitude).
The simulation is initialised with a potential field extrapolation based on a synoptic magnetogram for Carrington Rotation CR1910 from US National Solar Observatory, Kitt Peak (NSO/KP), corrected for differential rotation to represent 1996 May 15.
The coronal magnetic field is then evolved continuously, driven by the emergence of new active regions and shearing by large-scale photospheric motions. The latter are imposed on the lower boundary $r=R_\odot$ through the standard surface flux transport model \citep{leighton1964,sheeley2005}. Here we use a supergranular diffusivity $D=450{\rm km}^2{\rm s}^{-1}$, the \citet{snodgrass1983} differential rotation profile, and the meridional flow profile of \citet{schuessler2006} with peak speed $11\,{\rm ms}^{-1}$. In order for the simulation to reproduce the low polar field strengths observed after reversal in Cycle 23, we reduce the active region tilt angles by 20\% \citep{cameron2010,jiang2011}, and include an additional decay term for the photospheric magnetic field with a timescale of $10$ years \citep{schrijver2002,schrijver2008}.

Active regions are inserted in the form of idealised three-dimensional magnetic bipoles \citep{yeates2008}, whose properties (location, size, magnetic flux and tilt angle) are determined from NSO/KP synoptic magnetograms. Until 2003 these were taken with the older vacuum telescope, and from 2003 onwards with SOLIS (Synoptic Optical Long-Term Investigations of the Sun). For the present simulations we do not insert any bipoles to replace those missed during the three data gaps (CR2015-16, CR2040-41, CR2091). In total, between CR1911 (1996 June) and CR2110 (2011 May), we insert 1838 bipoles. The regions vary in flux content from $2\times 10^{20}\,{\rm Mx}$ to $5.3\times 10^{22}\,{\rm Mx}$.

As stated above, we reduce the tilt angles by 20\% to help reproduce the low polar fields observed after reversal in Cycle 23. A linear fit to the reduced tilt angles gives $\sin({\rm tilt})=0.38\sin(\lambda_0)-0.005$, 
which reduces the slope of the relationship compared to the magnetogram observations of either \citet{wang1989} (who used NSO/KP data for Cycle 21) or \citet{stenflo2012} (who used SOHO/MDI magnetograms). However, it is still greater than found in the white light study of \citet{dasiespuig2010} for Cycles 15 to 21. In principle, the same resulting polar fields could be obtained with the original bipole tilt angles by instead invoking a more complex meridional flow profile \citep{jiang2010}.

The three-dimensional bipoles have a twist parameter $\beta$ describing emergence of magnetic helicity within the bipole \citep{yeates2008}. Since the vector magnetic observations required to determine $\beta$ are not available for every active region in Cycle 23, we select $\beta$ statistically for each new bipole \citep{yeates2010}. Although we have shown that this emerging helicity affects the chirality of $\sim 32\%$ of filaments at lower latitudes \citep{yeates2009}, these are mainly those located within recently-emerged active regions. After only 30--50 days, differential rotation tends to reverse the chirality of PILs inside any bipoles inserted with the minority sign of $\beta$ \citep{yeatesthesis}. Thus the choice of $\beta$ distribution has little consequence for the higher latitudes considered here. We have verified that the same high latitude chirality pattern is recovered if $\beta=0$ for all bipoles.

\section{Results}

\begin{figure}
\includegraphics[width=0.8\textwidth]{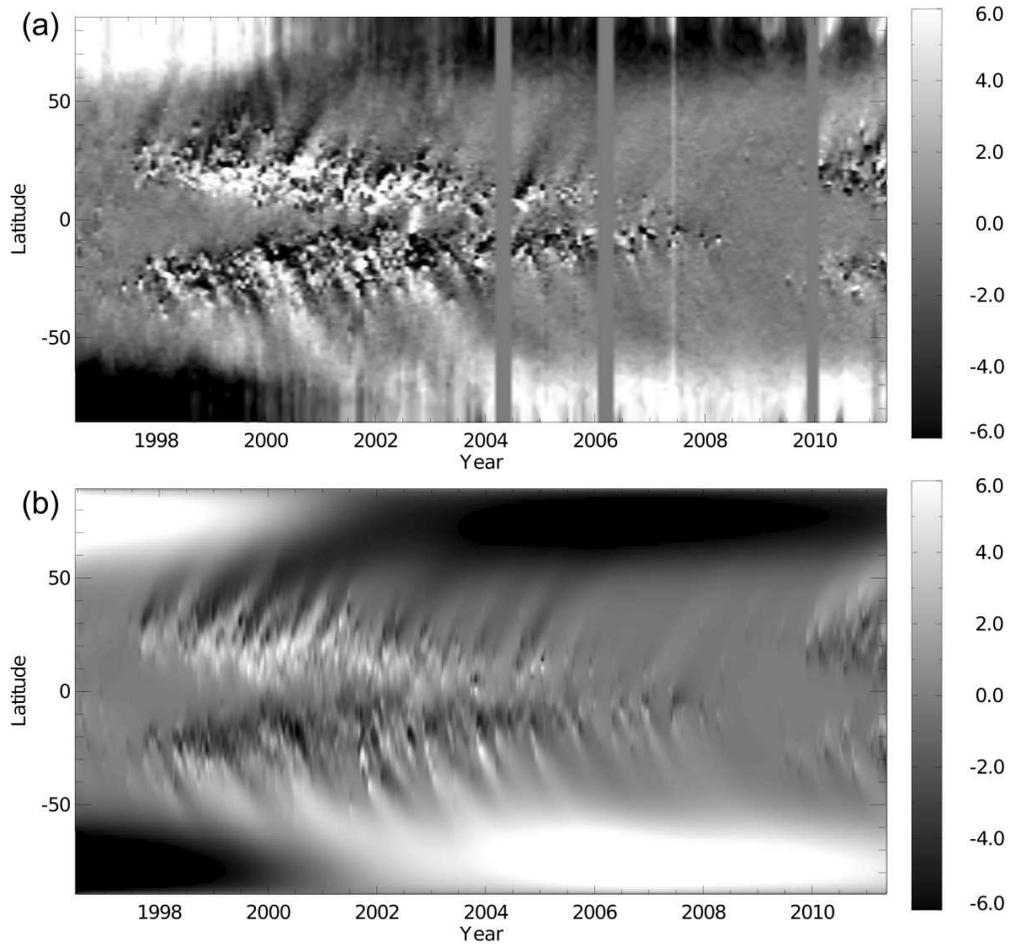}
\caption{Butterfly diagrams showing longitude-averaged radial magnetic field $B_{0r}$ over time, in (a) NSO/KP synoptic magnetograms and (b) the simulation. Three data gaps are evident in (a).
\label{fig2}}
\end{figure}

Figure \ref{fig2} shows that the photospheric radial magnetic field $B_{0r}$ in the observed magnetograms (Figure \ref{fig2}a) is generally well reproduced by the simulation (Figure \ref{fig2}b).

\begin{figure}
\includegraphics[width=\textwidth]{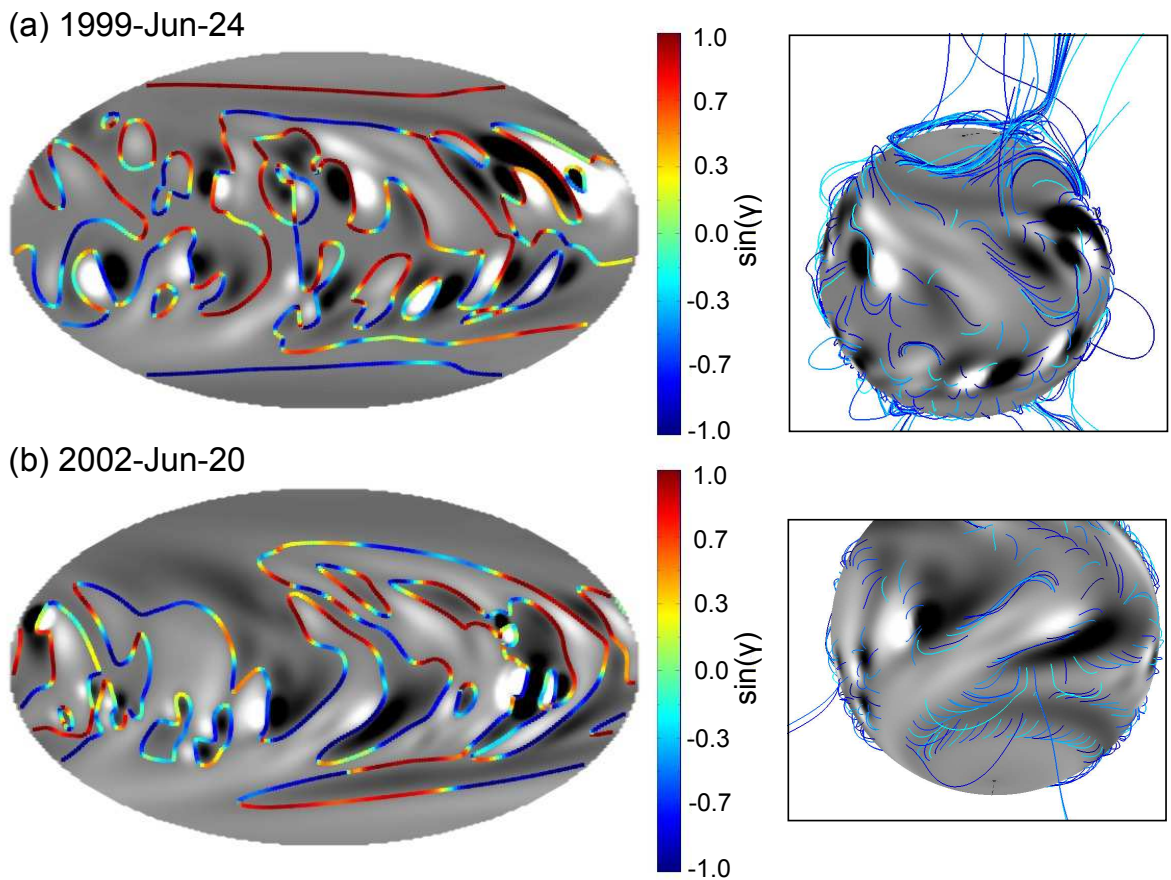}
\caption{Snapshots of the Cycle 23 non-potential simulation (a) before and (b) after polar field reversal. Left images show photospheric PILs colored according to skew $\sin\gamma$ at $r=1.033R_\odot$ (in Mollweide equal-area projection), while right images show selected coronal field lines traced from this height. Dextral skew corresponds to $\sin\gamma > 0$ (red) and sinistral to $\sin\gamma < 0$ (blue). In both cases, grey shading shows $B_{0r}$ on the photosphere $R_\odot$ (white positive, black negative; saturation $\pm 30\,{\rm G}$). The left images are also available as an mpeg
animation in the electronic edition of the Journal. 
\label{fig3}}
\end{figure}

To analyse the chirality, snapshots of the global three-dimensional magnetic field were stored every 14 days during the simulation. For each snapshot, PILs were identified automatically on the photosphere $r=R_\odot$, along with the overlying chirality of the coronal magnetic field at height $r=r_1$. We take $r_1=1.033R_\odot$ to approximate the height of filaments. Now denote the horizontal field at $r=r_1$ by ${\bf B}_1^\perp$, and the radial field in the photosphere by $B_{0r}$. We characterise the chirality at height $r=r_1$ by the ``skew''
\begin{equation}
\sin\gamma = \frac{{\bf e}_r\cdot\nabla B_{0r}\times{\bf B}_1^\perp}{|\nabla B_{0r}||{\bf B}_1^\perp|},
\end{equation}
so that $\sin\gamma<0$ corresponds to sinistral skew and $\sin\gamma>0$ to dextral skew. Two example snapshots from the simulation, on 1999 June 24 and 2002 June 20, are shown in Figure \ref{fig3}. The color shading along photospheric PILs indicates the skew calculated above them.

\begin{figure}
\includegraphics[width=0.8\textwidth]{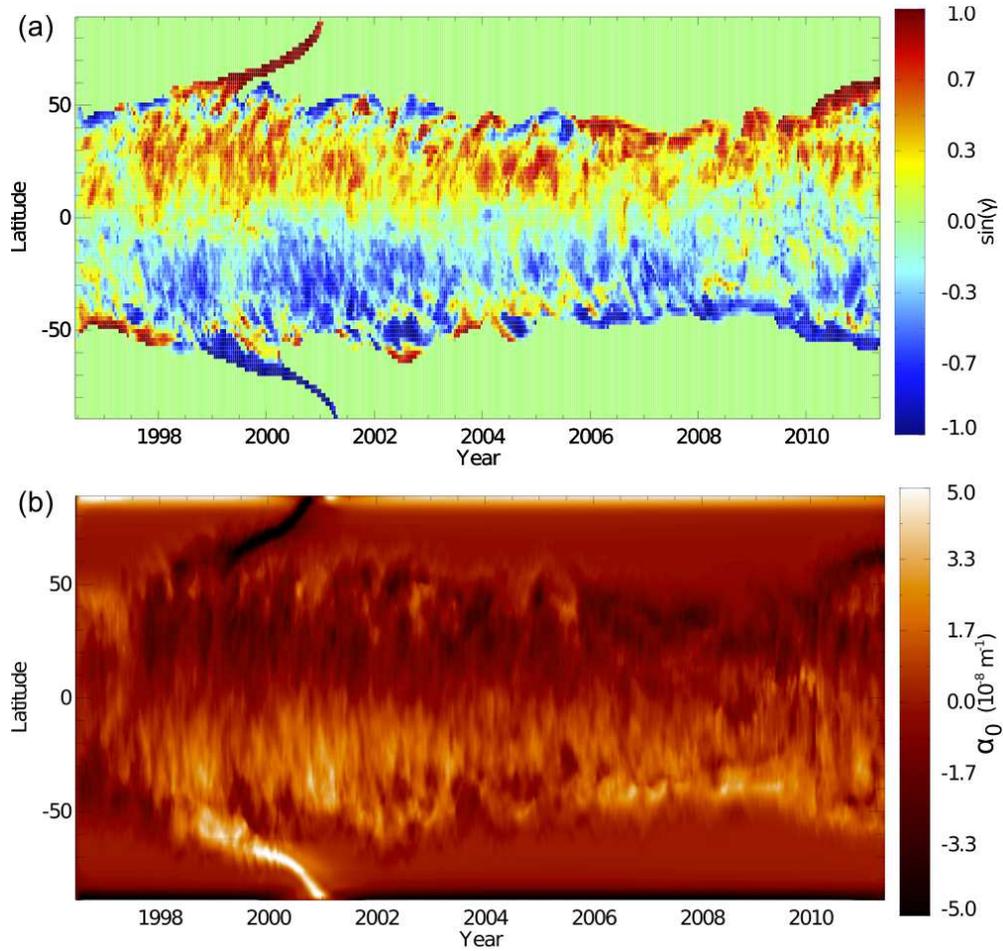}
\caption{Butterfly diagrams showing longitude-averages of (a) skew $\sin\gamma$, and (b) current helicity density $\alpha_0$, during the Cycle 23 non-potential simulation. Both are measured at height $r=1.033R_\odot$.
See the electronic edition of the Journal for a color version of this figure.
\label{fig4}}
\end{figure}

Figure \ref{fig4}(a) shows the longitude-averaged skew angle along PILs in 95 latitude bins, as a function of time through the simulation. Several features are apparent:
\begin{enumerate}
\item In and around active regions the majority pattern of chirality predominates, i.e., dextral in the northern hemisphere and sinistral in the southern hemisphere, although this is an overall mean pattern with significant local fluctuations in both magnitude and sign (as seen in Figure \ref{fig3}). The origin of this pattern, including the exceptions, is explained by \citet{yeates2009} \citep[see also][]{yeatesthesis}.
\item During the period of few active regions from 2007 to 2010, there is more mixed chirality at lower latitudes.
\item Until 1998, and during the declining phase from 2001 to 2006, there is a tendency for minority chirality on the high-latitude PILs (sinistral in the north, dextral in the south).
\item During the ``rush-to-the-poles'' between 1998 and mid-1999, the polar crowns exhibit the majority chirality pattern.
\item From 2006 onward, the majority chirality dominates at high latitudes once more, continuing into the start of Cycle 24.
\end{enumerate}

For comparison, \ref{fig4}(c) shows the longitude-averaged current helicity density $\alpha_0$ \citep[Equation \ref{eqn:alpha0}; see also][]{yeates2008a}, at the same height $r=1.033R_\odot$. As expected, the sign of $\alpha_0$ correlates approximately with that of the skew angle, where positive $\alpha_0$ is associated with negative skew angle, and {\it vice versa}. (The concentrations of $\alpha_0$ at the poles are due to twisting up of the vertical field there by the differential rotation.)


\section{Discussion}

The chirality pattern at high latitudes in Figure \ref{fig4}(b) differs from the earlier 6-month simulations \citep{yeates2008}. In those simulations, high-latitude filaments remained predominantly sinistral in the northern hemisphere and dextral in the southern hemisphere: opposite to the majority pattern. In the new 15-year simulation, the majority pattern is found in two independent epochs: once during the rising phase up to polar reversal (including the period of the previous simulations), and again from the middle of the declining phase of Cycle 23. We argue that:
\begin{enumerate}
\item Transport of helicity from active latitudes is responsible for overcoming the effect of differential rotation on the polar crown over much (but not all) of the cycle.
\item The initial period of opposite chirality (before 1999) is an artefact of the initial condition of the simulation, indicating the importance of magnetic memory in the corona. Note that such a period of opposite chirality does not recur at the start of Cycle 24 (from 2009).
\end{enumerate}

An example where differential rotation clearly wins over helicity transport is shown in Figure \ref{fig3}(b), during the declining phase of the solar cycle. Here, minority dextral chirality is found on the poleward arm of a switchback in the southern hemisphere. The field lines (Figure \ref{fig3}, bottom right) show a sheared arcade, long disconnected from active region flux. The tendency for minority chirality on such switchbacks in the declining but not in the rising phase matches the predictions made from single-bipole simulations by \citet{mackay2001}.

\begin{figure}
\includegraphics[width=0.8\textwidth]{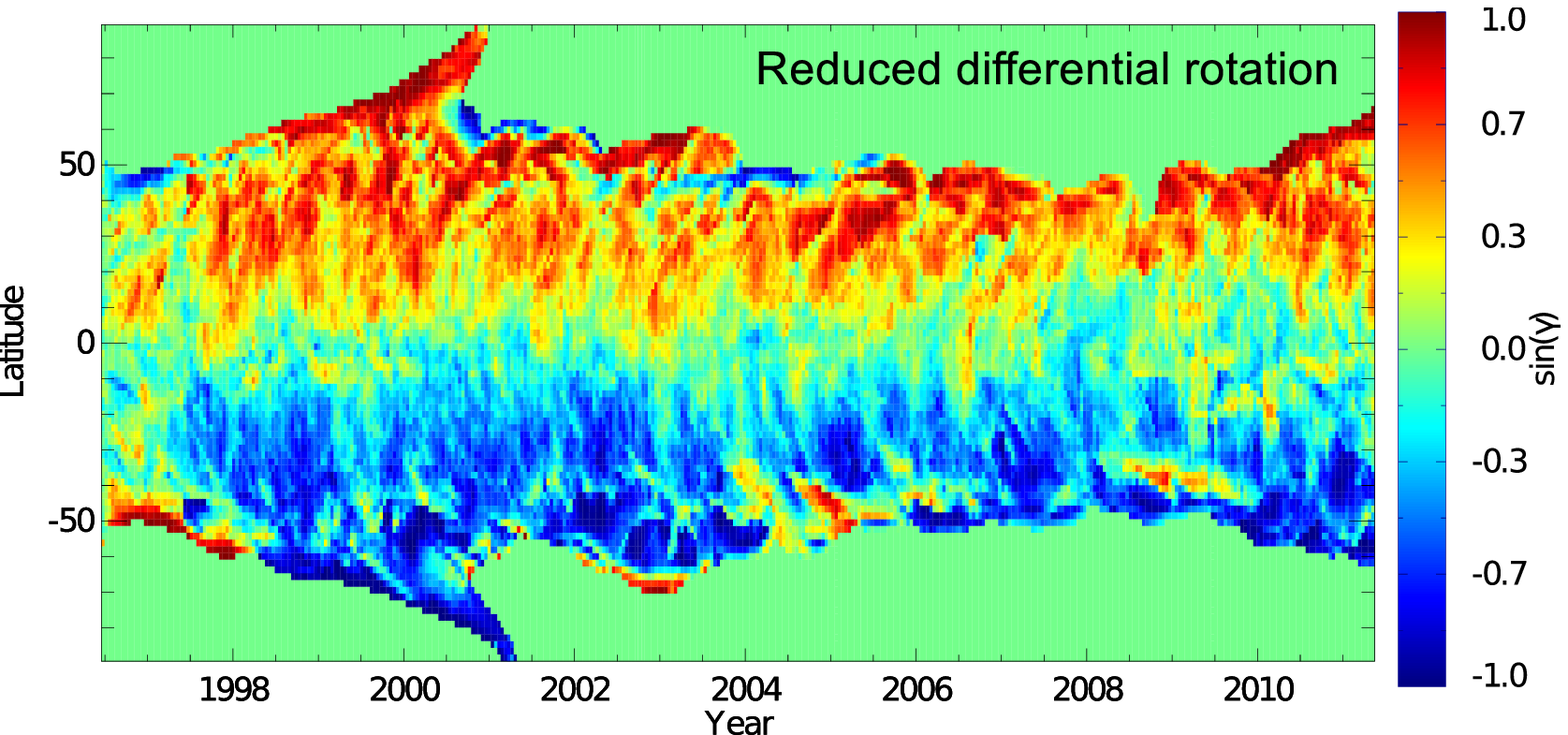}
\caption{Longitude-averaged skew $\sin\gamma$ at height $r=1.033R_\odot$ for the simulation with reduced differential rotation, showing reduced areas of minority chirality. \citep[Note that the longitude-averaged $B_{0r}$ in the photosphere is not affected by changing the differential rotation;][]{leighton1964}. See the electronic edition of the Journal for a color version of this figure.
\label{fig5}}
\end{figure}

By constrast, the north polar crown during the rising phase prior to polar reversal (Figure \ref{fig3}a) exhibits a strong flux rope with majority chirality, with field lines directly connecting to lower latitudes. To further demonstrate that the polar crown chirality is controlled by a balance between differential rotation and helicity transport, we repeated the simulation with the latitudinal differential gradient reduced to half of its original slope. This leads to less minority chirality at high latitudes (Figure \ref{fig5}), consistent with the greater dominance of helicity transport. Even in the original simulation, majority chirality takes over at all latitudes later in the cycle (from 2006 onward), because the highest-latitude PILs at this time are nearer to the equator than those earlier in the cycle, so that helicity from lower latitudes can more easily dominate.

The initial period of minority chirality at high latitudes does not recur at the start of Cycle 24 in the simulation. This suggests it may be an artefact of the initial potential field. Majority chirality on the polar crowns during the minimum and rising phase of the solar cycle would be required in order to be consistent with the reported polarization measurements \citep{leroy1978,leroy1983}. Indeed, when we simulate two consecutive cycles with emerging active regions chosen at random from statistical distributions (not illustrated here), we find this minority chirality only at the start of the simulation, not at the start of subsequent cycles.
Because the high-latitude field depends on transport of helicity from lower latitudes, differential rotation dominates early in the simulation, producing minority chirality on the polar crowns. In effect, the high latitude field has a finite memory of around 12 months for the pre-existing magnetic topology at the start of the simulation. This explains why the previous 6-month simulations \citep{yeates2008} found minority chirality on the polar crowns in 1999.

In conclusion, the NP model makes robust predictions for the chirality of high-latitude filaments: majority chirality over much of the cycle (including during polar field reversal), but more mixed chirality between the polar reversal and middle of the declining phase. The first prediction matches the available polarization measurements of prominences from earlier solar cycles, but further measurements are needed to confirm the second prediction. Mixed chirality is also predicted at lower latitudes during particularly quiet periods such as the recent Cycle 23 minimum.
There are some indications that the hemispheric pattern of active region helicity was also weaker during the declining phase of Cycle 23 \citep{tiwari2009}, although this remains statistically uncertain \citep{pevtsov2008}. However, the multitude of factors affecting filament chirality mean that we do not expect to see the same time variations in the filament and active region hemispheric patterns \citep{yeatesthesis,yeates2009}. Testing our predictions against prominence measurements will best validate the NP model for the global coronal magnetic field.
Of course, our model at present considers only the large-scale magnetic field within which filaments form. The smaller scale magnetic field structure inside filaments/prominences themselves is inferred to be rather complex and is the subject of ongoing research \citep{merenda2006,vanballegooijen2010}.







\acknowledgments

DHM acknowledges financial support from a Leverhulme Trust Research Grant. Research leading to these results has received funding from the European Commission's Seventh Framework Programme (FP7/2007-2013) under grant agreement SWIFF (263340). Simulations used the STFC and SRIF funded UKMHD  cluster at the University of St Andrews. SOLIS data used here are produced cooperatively by NSF/NOAO and NASA/LWS.

\clearpage


\end{document}